\documentclass[prd,aps,floats]{revtex4}

\usepackage{graphicx,xcolor}
\usepackage{epsfig}
\usepackage{amssymb}
\usepackage{amsmath,amssymb,textcomp,array}

\newcommand{\mn}{{Mon.\@ Not.\@ Roy.\@ Ast.\@ Soc.\ }}
\newcommand{\asta}{{Astron.\@ Astrophys.\ }}
\newcommand{\aj}{{Astron.\@ J.\ }}

\newcommand{\plb}{{Phys.\@ Lett.\@ B\ }}

\newcommand{\jcap}{J.~Cosmol.~Astropart.~Phys.~}

\newcommand{\etal}{{et al.,~}}
\newcommand{\ie}{{i.e.,~}}
\newcommand{\etc}{{e.t.c}}

\newcommand{\eg}{{e.g.,~}}

\newcommand{\beq}{\begin{equation}}
\newcommand{\eeq}{\end{equation}}
\newcommand{\ber}{\begin{eqnarray}}
\newcommand{\eer}{\end{eqnarray}}
\newcommand{\lleq}{\lower0.9ex\hbox{ $\buildrel < \over \sim$} ~}
\newcommand{\ggeq}{\lower0.9ex\hbox{ $\buildrel > \over \sim$} ~}
\newcommand{\lsim}{\ \lower-1.5pt\vbox{\hbox{\rlap{$<$}\lower5.3pt\vbox{\hbox{$\sim$}}}}\ }
\newcommand{\gsim}{\ \lower-1.5pt\vbox{\hbox{\rlap{$>$}\lower5.3pt\vbox{\hbox{$\sim$}}}}\ }

\newcommand{\omt}{\Omega_{0 m}}
\newcommand{\omr}{\Omega_{0 r}}
\newcommand{\omg}{\Omega_{0 \gamma}}

\newcommand{\omb}{\Omega_{0b}}
\newcommand{\omk}{\Omega_{\kappa}}
\newcommand{\oml}{\Omega_{\ell}}
\newcommand{\omsig}{\Omega_{\sigma}}
\newcommand{\omlb}{\Omega_{\Lambda_b}}
\newcommand{\omdr}{\Omega_{C}}

\newcommand{\lya}{Ly$\alpha$~}
\newcommand{\la}{{l}_A}

\begin{document}

\title{Constraining the Cosmology of the Phantom Brane using Distance Measures}

\author{Ujjaini Alam$^{[a]}$, Satadru Bag$^{[b]}$, Varun Sahni$^{[b]}$}
\affiliation{$^{[a]}$Physics \& Applied Mathematics Unit, Indian Statistical Institute, Kolkata India}
\email{ujjaini.alam@gmail.com}
\affiliation{$^{[b]}$Inter University Center for Astronomy \& Astrophysics, Pune India}
\email{satadru@iucaa.in,varun@iucaa.in}

\thispagestyle{empty}

\sloppy

\begin{abstract}
  \small{
The phantom brane has several important distinctive features: (i) Its
equation of state is phantom-like, but there is no future `big rip'
singularity, (ii) the effective cosmological constant on the brane is
dynamically screened, because of which the expansion rate is {\em
  smaller} than that in $\Lambda$CDM at high redshifts. In this paper,
we constrain the Phantom braneworld using distance measures such as
Type Ia supernovae (SNeIa), Baryon Acoustic Oscillations (BAO), and
the compressed Cosmic Microwave Background (CMB) data. We find that
the simplest braneworld models provide a good fit to the data. For
instance, BAO +SNeIa data can be accommodated by the braneworld for a
large region in parameter space $0 \leq \oml \lleq 0.3$ at
$1\sigma$. The Hubble parameter can be as high as $H_0 \lleq 78 \ {\rm
  km} \ {\rm s}^{-1} \ {\rm Mpc}^{-1}$, and the effective equation of
state at present can show phantom-like behaviour with $w_0 \lleq -1.2$
at $1\sigma$. We note a correlation between $H_0$ and $w_0$, with
higher values of $H_0$ leading to a lower, and more phantom-like,
value of $w_0$. Inclusion of CMB data provides tighter constraints
$\oml \lleq 0.1$. (Here $\oml$ encodes the ratio of the five and four
dimensional Planck mass.) The Hubble parameter in this case is more
tightly constrained to $H_0 \lleq 71 \ {\rm km} \ {\rm s}^{-1} \ {\rm
  Mpc}^{-1}$, and the effective equation of state to $w_0 \lleq -1.1$.
Interestingly, we find that the universe is allowed be closed or open,
with $-0.5 \lleq \omk \lleq 0.5$, even on including the compressed CMB
data. There appears to be some tension in the low and high $z$ BAO
data which may either be resolved by future data, or act as a pointer
to interesting new cosmology.}
\end{abstract}

\maketitle

\section{Introduction}\label{sec:intro}

The unexpected faintness of distant supernova Type Ia, as observed
concurrently by the Supernova Cosmology Project (SCP) and the High
Redshift Search Team (HZT) \cite{hzt,scp} in the late 1990s, has led
to the postulation of one of the most mystifying cosmological
phenomena-- the accelerated expansion of the Universe. One way to
explain this observational result is to theorize the existence of a
new form of energy, with negative pressure, often called `dark
energy'. Many different models have been suggested for this dark
energy, some of which are reviewed in \cite{de_rev,ss06}. Current
cosmological observations are commensurate with the cosmological
constant \cite{planck}, where the dark energy equation of state is
$-1$ and its energy density is constant. However, other dark energy
models are by no means ruled out \cite{sss14}, and the search for the
true nature of dark energy is a continuing process.

A different approach to the problem of cosmological acceleration
consists of introducing new physics in the gravitational
sector. Einsteinian gravity is very well tested within the solar
system, but may be modified on larger scales. Different models of
modified gravity have been suggested to explain the accelerated
expansion of the universe \cite{modgrav}, including $f(R)$ models,
galileons etc. We shall consider here a braneworld scenario, where our
observable universe is situated in a four-dimensional brane embedded
in a fifth dimension, the `bulk', and the accelerated expansion of the
universe is a consequence of this modification of gravity. Braneworld
scenarios could have important cosmological consequences.  For
instance, (i) the Randall-Sundrum (RS) model \cite{rs1}, which
modifies gravity at small scales, could potentially explain the galaxy
rotation curves in lieu of dark matter \cite{rs2}, (ii) An RS-type
braneworld, but with a time-like extra dimension, makes the universe
bounce at early times, alleviating thereby the big bang singularity
\cite{rs3}.  The braneworld models which produce accelerated expansion
of the universe tend to modify gravity on large scales. An early
example, the DGP model, was constructed in \cite{dgp} while a more
general braneworld model containing the induced gravity term as well
as cosmological constants in the bulk and on the brane, has been
studied in \cite{brane1, brane2, brane3, brane}.

In this work we study a braneworld model for the accelerated expansion
of the universe that was introduced in \cite{brane} and discussed in
greater detail in \cite{loiter,brane4}.  We revisit this model in the
context of observations of the cosmological distance and attempt to
constrain it from the latest data. In the following sections, we first
define our braneworld model in section~\ref{sec:brane}, discuss the
data and methodology in section~\ref{sec:data}, show the results of
our analysis in section~\ref{sec:res}, and present our conclusions in
section~\ref{sec:concl}.

\section{Cosmological Evolution of the Braneworld model}\label{sec:brane}

We consider a braneworld scenario where the equations of motion are
derived from the action \cite{brane}
\ber \label{eq:action} 
S = M^3 \left[\int_{\rm bulk} \left( R_5 - 2 \Lambda_{b} \right) - 2
  \int_{\rm brane} K \right]
+ \int_{\rm brane} \left( m^2 R_4 - 2 \sigma \right)&& \nonumber\\ 
+ \int_{\rm
  brane} L \left( h_{\alpha\beta}, \phi \right)&& \,\,,  
\eer
where, $R_5$ is the scalar curvature of the metric $g_{ab}$ in the
five-dimensional bulk, and $R_4$ is the scalar curvature of the
induced metric $h_{\alpha\beta}$ on the brane. The quantity $K =
K_{\alpha\beta} h^{\alpha\beta}$ is the trace of the extrinsic
curvature $K_{\alpha\beta}$ on the brane defined with respect to its
inner normal. $L (h_{\alpha\beta}, \phi)$ is the four-dimensional
matter field Lagrangian, $M$ and $m$ denote, respectively, the
five-dimensional and four-dimensional Planck masses, $\Lambda_{ b}$ is
the bulk cosmological constant, and $\sigma$ is the brane tension.
Integrations in Eq~(\ref{eq:action}) are performed with respect to the
natural volume elements on the bulk and brane.  The presence of the
brane curvature term $m^2\int_{\rm brane}R_4$ in Eq~(\ref{eq:action})
introduces the length scale $\ell = 2m^2/M^3$. On short length scales
$r \ll {\ell}$ (early times) one recovers general relativity, whereas
on large length scales $r \gg {\ell}$ (late times) brane-specific
effects begin to play an important role, leading to the acceleration
of the universe at late times.

The cosmological evolution of the braneworld is described by the
Hubble parameter 
\beq \label{eq:hubble_ev} 
\hspace{-1.2cm}H^2 + \frac{\kappa}{a^2} = \frac{\rho + \sigma}{3 m^2} + \underline{\frac{2}{{\ell}^2} \left[1 \pm \sqrt{1 + {\ell}^2
      \left(\frac{\rho + \sigma}{3 m^2} - \frac{\Lambda_{ b}}{6} - \frac{C}{a^4} \right)}~ \right]} \,, ~ ~~ m^2 = \frac{1}{8\pi G},  
\eeq
where $H = \dot{a}/a$ is the Hubble parameter, $\rho=\rho(t)$ is the
energy density of matter and radiation on the brane, $C/a^4$
represents the dark radiation term and $\kappa$, the curvature of the
universe.  The underlined terms make the braneworld models different
from standard FLRW cosmology. The `$\pm$' signs in
Eq~(\ref{eq:hubble_ev}) correspond to the two separate ways in which
the brane can be embedded in the higher dimensional bulk. The two
signs represent two branches of cosmological solutions, the `$+$' sign
denoting the `self-accelerating' branch which can model late-time
acceleration without cosmological constant in the bulk or on the
brane, while the `$-$' sign represents the `normal' branch where at
least a brane tension is required to accelerate the expansion. It has
been shown that the self-accelerating branch is plagued by ghost
instability issues at least in the DGP model of braneworlds
\cite{ghost}. In this paper, we limit ourselves to the the `$-$' sign,
or the normal branch, which exhibits phantom-like behaviour. A version
of this model has been previously studied in context of an older
dataset in \cite{brane_data}, and we now extend this analysis for the
newest data using all the different braneworld parameters.

The reduced Hubble parameter $h(z)=H(z)/H_0$ can be calculated from
(\ref{eq:hubble_ev}) to be
\ber\label{eq:hubble}
h^2(z) &=&  \omr(1+z)^4+\omt(1+z)^3+\Omega_\kappa(1+z)^2+
\omsig+2\oml \\\nonumber &&-2\sqrt{\oml}\sqrt{\omr(1+z)^4+\omt(1+z)^3+\omsig+\oml+\omlb+
\omdr(1+z)^4}\,\,,
\eer 
with the additional constraint relation
\beq\label{eq:omsig}
\omsig=1-\omr-\omt-\Omega_\kappa+2\sqrt{\oml}\sqrt{1+\omlb+\omdr-\Omega_\kappa}\,\,.
\eeq
Here
\beq \label{eq:omegas}
\Omega_{0m} =  {\rho_{0m} \over 3 m^2 H_0^2} \, , \Omega_{0r} =  {\rho_{0r} \over 3 m^2 H_0^2} \, , \Omega_\kappa = -
{\kappa \over a_0^2 H_0^2} \, , \Omega_\sigma = {\sigma \over 3 m^2
H_0^2} \, , \Omega_{\ell} = {1 \over {\ell}^2 H_0^2} \, ,
\omlb = - {\Lambda_{b} \over 6 H_0^2} \, , \omdr = -\frac{C}{a_0^4H_0^2}
\eeq
are dimensionless parameters. In the limit $\oml \rightarrow 0$, the
braneworld reduces to the $\Lambda$CDM model.  The parameters to be
constrained are $\omt, \ \oml, \ \omlb, \Omega_\kappa, \ \omdr$ and
$H_0$ ($\omsig$ is constrained by Eq~(\ref{eq:omsig})). The value of
the radiation density can be calculated from the CMB temperature or
from BBN considerations to a high degree of accuracy.

\begin{figure}
\begin{center}
\includegraphics[width=0.5\textwidth]{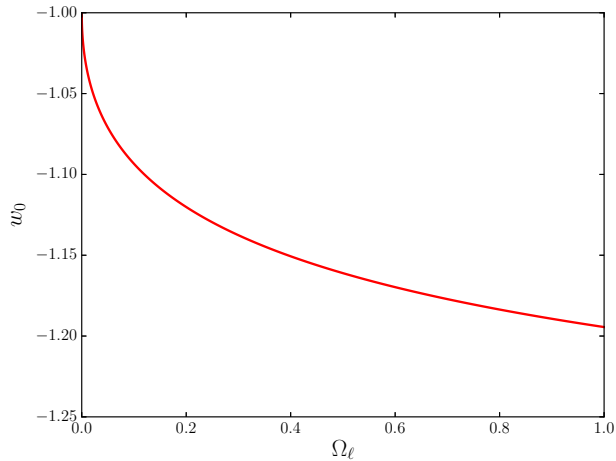}  \\ [0.0cm]
\caption{\footnotesize
The current value of the effective equation of state of dark energy
($w_0$) in the braneworld model (\ref{eq:hubble0}) is shown as a
function of $\Omega_{\ell}$ ($\Omega_{0m}=0.28$ is assumed). For $\oml
\rightarrow 0$, one recovers $\Lambda$CDM limit.}
\label{fig:pole_w0}
\end{center}
\end{figure}

A simpler variant of the above model is obtained by setting
$\Omega_\kappa = \omdr = \omlb = 0$ and neglecting the presence of
radiation at low redshifts.  In this case (\ref{eq:hubble}) \&
(\ref{eq:omsig}) reduce to
\beq\label{eq:hubble00}
h^2(z) =  \omt(1+z)^3+\omsig+2\oml -2\sqrt{\oml}\sqrt{\omt(1+z)^3+\omsig+\oml}\,\,,
\eeq
\beq\label{eq:omsig00}
\omsig=1-\omt+2\sqrt{\oml}\,\,.
\eeq

This model has several interesting features which hold for the entire 
normal-branch Braneworld family. 

\begin{enumerate}

\item
First and foremost is the fact that the current value of the effective
equation of state is phantom-like, \ie $w_{\rm eff} < -1$. To
appreciate this let us define the energy density and pressure of dark
energy on the brane as follows \cite{ss06}
\ber\label{eq:energy}
\rho_{\rm DE} &=&
\frac{3H^2}{8\pi G}(1 - \Omega_{m})\nonumber\\
p_{\rm DE} &=& \frac{H^2}{4\pi G}(q - \frac{1}{2})~,
\eer
where 
\beq\label{eq:deceleration}
q\equiv -\ddot a/aH^2 = x\frac{H'(x)}{H(x)} - 1 \, , ~~~ x = 1 + z \,,
\eeq
is the deceleration parameter (the prime denotes differentiation with
respect to $x$ or $z$) and $\Omega_m$ is the total density of
non-relativistic matter in terms of its critical value
\beq\label{eq:omz}
\Omega_{m}(z) = \frac{\omt(1+z)^3}{h^2(z)}~.
\eeq
The effective equation of state (EOS) of dark energy, $w_{\rm eff} =
p_{\rm DE}/\rho_{\rm DE}$, is then given by
\beq\label{eq:state}
w_{\rm eff}(z) = {2 q(z) - 1 \over 3 \left( 1 - \Omega_{m}(z) \right)}
\eeq

Substituting from \eqref{eq:deceleration}, \eqref{eq:omz} \&
\eqref{eq:hubble00} into (\ref{eq:state}) we get $w_{\rm eff}(z)$ for
the Phantom braneworld as
\beq\label{eq:w_bw}
w_{\rm eff}(z)=-1-\frac{\Omega_m(z)}{1-\Omega_m(z)}\sqrt{\frac{\oml}{\omt(1+z)^3+\omsig+\oml}}\;.
\eeq
At the present epoch ($z=0$),
\beq\label{eq:weff_bw_0}
w_0 \equiv w_{\rm eff}(z=0)=-1-\dfrac{\Omega_{0m}}{1-\Omega_{0m}} \left(\dfrac{\sqrt{\Omega_{\ell}}}{1+\sqrt{\Omega_{\ell}}}\right)\;,
\eeq
demonstrating that the present value of the effective equation of
state of the dark energy is {\em phantom-like}, i.e. $w_0 < -1$.
Figure \ref{fig:pole_w0} shows $w_0$ as a function of $\Omega_\ell$.
We find that $w_0 \rightarrow -1/(1-\Omega_{0m})$ asymptotically, as
$\Omega_{\ell} \rightarrow \infty$.

\begin{figure*}
\begin{center}
\includegraphics[width=0.5\textwidth]{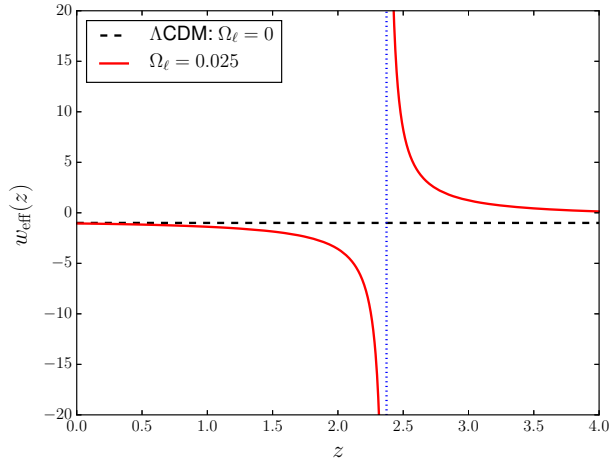}  \\ [0.0cm]
\caption{\footnotesize
The effective equation of state of dark energy ($w_{\rm eff}$) is
shown as a function of redshift for $\Omega_{\ell}=0.025$, assuming
$\omt=0.28$.  A pole occurs at $z_p \approx 2.372$. At large redshift,
$w_{\rm eff}(z) \rightarrow -1/2$ for any non zero value of $\oml$ and
$\omt$. For $\oml=0$, i.e. in the $\Lambda$CDM limit, the pole
disappears as shown by the dashed line.}
\label{fig:pole_ol25}
\end{center}
\end{figure*}

\begin{figure*}
\begin{center}
\includegraphics[width=0.5\textwidth]{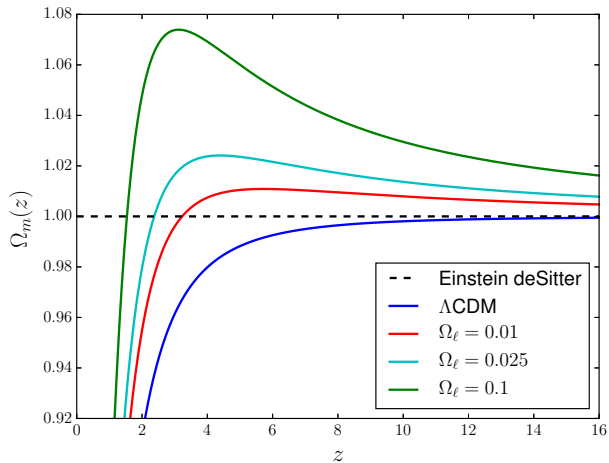}  \\ [0.0cm]
\caption{\footnotesize
$\Omega_m(z)$, given by \eqref{eq:omz}, is plotted against the
redshift $z$ for various $\oml$, assuming $\omt=0.28$. During matter
domination (large $z$), $\Omega_m(z)$ approaches unity. In the Phantom
braneworld $\Omega_m(z)$ possesses a maximum and $\Omega_m(z)>1$ while
$z>z_p$. Pole in $w_{\rm eff}(z)$ occurs at $z=z_p$ when
$\Omega_m(z_p)=1$. As $\oml$ increases, $\Omega_m(z)$ becomes unity at
lower redshift, i.e. $z_p$ decreases, which is explicitly shown in
figure \ref{fig:pole_zp_all}. For Einstein deSitter universe
$\Omega_m(z)=1$ always. }
\label{fig:Omz}
\end{center}
\end{figure*}

\begin{figure*}
\begin{center}
\includegraphics[width=0.5\textwidth]{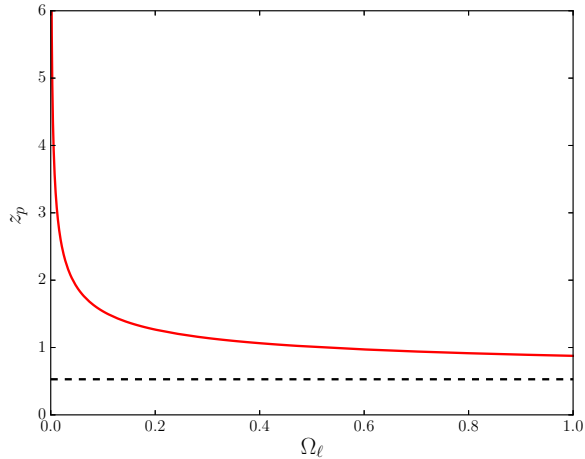}  \\ [0.0cm]
\caption{\footnotesize
The redshift of the pole in $w_{\rm eff}(z)$ is shown as a function of
$\Omega_{\ell}$.  The dashed line corresponds to the asymptotic value
of $z_p$: $z_p \rightarrow \left[\left(1/\omt\right)^{1/3}-1\right]$
for $\Omega_\ell \to \infty$.  For $\Omega_{0m}=0.28$ the asymptote is
at $z_p\approx 0.53$.}
\label{fig:pole_zp_all}
\end{center}
\end{figure*}

\item
A second important feature of the Phantom brane is that the effective
cosmological constant on the brane can be {\em screened}. This can
easily be seen by rewriting (\ref{eq:hubble00}) in the more suggestive
form
\beq
h^2(z) =  \omt(1+z)^3 + \Omega_\Lambda - f(z)
\eeq
where $\Omega_\Lambda = \omsig+2\oml$ and $f(z)$ is the screening term
$f(z) = -2\sqrt{\oml}\sqrt{\omt(1+z)^3+\omsig+\oml}$ whose value {\em
  increases} with redshift.  The presence of this term permits the
expansion rate to {\em fall below} the $\Lambda$CDM value of $h^2(z) =
\omt(1+z)^3 + \Omega_\Lambda$ at high redshifts
\cite{brane,Lue:2004za,sss14}. The screening mechanism, operational in
the braneworld$^1$ \footnotetext[1]{The cosmological constant can also
  be dynamically screened in other cosmological scenario's, some of
  which are discussed in \cite{screened}.} can potentially be tested
by observations of $h(z)$. As pointed out in \cite{sss14}, the phantom
brane may provide a better fit to high-z BAO data than $\Lambda$CDM.
Future BAO data are likely to improve on this result by providing very
accurate measurements of the expansion history of the
universe. Combining observations of $h(z)$ with the Om diagnostic
\cite{omz,sss14}, and eventually with the Statefinder \cite{state},
would allow one to assess the nature of dark energy in a model
independent manner.

As noted in \cite{sss14}, a key feature of screened dark energy models
is that if $f(z)$ increases monotonically with redshift, then
eventually the cosmological constant, $\Omega_\Lambda$, will be
cancelled by $f(z)$, so that $h^2(z_p) = \omt(1+z_p)^3$. At this
redshift, $z_p$, the effective equation of state of dark energy will
develop a pole at which $w_{\rm eff}(z_p) \to \infty$
\cite{loiter,sss14}.  In the context of the Phantom brane, the pole in
$w_{\rm eff}(z)$ is shown in figure \ref{fig:pole_ol25}.  It is easy
to see that the presence of the pole is generic and arises when
$\Omega_m(z_p) = 1$ in the denominator of \eqref{eq:w_bw}. Actually
$\Omega_m(z)$ in the Phantom braneworld possesses a maximum and
remains greater than unity for $z>z_p$, as shown in figure
\ref{fig:Omz}.  This figure informs us that, for increasing values of
$\oml$, $\Omega_m(z)$ reaches unity at lower redshifts. This implies
that $z_p$ decreases with increasing $\oml$.  The redshift of the
pole, $z_p$, is given by
\beq
(1+z_p)^3 = \frac{\omsig^2}{4\Omega_{0m}\Omega_{\ell}}~.
\eeq
The value of $z_p$ is plotted against $\Omega_{\ell}$ in figure
\ref{fig:pole_zp_all}. Using the closure relation \eqref{eq:omsig0},
we find that $(1+z_p)^3 \rightarrow 1/\omt$ asymptotically as $\oml
\rightarrow \infty$.  The presence of a pole in the EOS of dark energy
therefore emerges as a {\em smoking gun} test for this class of
Braneworld models. We note that such a pole may also be present for
other dark energy models in which the dark energy density crosses
zero.

\end{enumerate}

The above characteristics for this subset of Phantom brane also hold
true for other subsets of this model which are considered in this
work.

\begin{itemize}

\item 
Our base braneworld model is a flat universe without dark radiation, \ie
$\kappa=0$, $C=0$. This is very similar to the the simplest variant
for the Phantom brane considered above, except that the radiation
density is explicitly considered as well, since high redshift data is
also considered. The reduced Hubble parameter has the form
\ber\label{eq:hubble0}
h^2(z) &=&  \omr(1+z)^4+\omt(1+z)^3+\omsig+2\oml \\\nonumber 
&&-2\sqrt{\oml}\sqrt{\omr(1+z)^4+\omt(1+z)^3+\omsig+\oml+\omlb}\,\,,
\eer 
with the additional constraint relation
\beq\label{eq:omsig0}
\omsig=1-\omr-\omt+2\sqrt{\oml}\sqrt{1+\omlb}\,\, .
\eeq

The effective equation of state at present is given by 
\beq
w_0 = -1 -\frac{1}{3}\frac{\sqrt{\oml}(4\omr+3\omt)}{(1-\omr-\omt)(\sqrt{1+\omlb}+\sqrt{\oml})}\,\,.
\eeq

The parameters to be fitted are $\omt, \ \oml, \ \omlb$ and $H_0$.

\item
We also study the Phantom brane including dark radiation as a
parameter, in a flat universe, \ie $\kappa=0, C \neq 0$.

The reduced Hubble parameter will therefore be given by
\ber\label{eq:hubble1}
h^2(z) &=&  \omr(1+z)^4+\omt(1+z)^3+\omsig+2\oml \\\nonumber 
&&-2\sqrt{\oml}\sqrt{\omr(1+z)^4+\omt(1+z)^3+\omdr(1+z)^4+\omsig+\oml+\omlb}\,\,,
\eer 
with the additional constraint relation
\beq\label{eq:omsig1}
\omsig=1-\omr-\omt+2\sqrt{\oml}\sqrt{1+\omlb+\omdr}\,\, .
\eeq

Here the effective equation of state at present takes the form 
\beq
w_0 = -1 -\frac{1}{3}\frac{\sqrt{\oml}(4\omr+3\omt+4\omdr)}{(1-\omr-\omt)(\sqrt{1+\omlb+\omdr}+\sqrt{\oml})}\,\,.
\eeq

The parameters to be fitted are $\omt, \ \oml, \ \omlb, \ \omdr$ and
$H_0$. The dark radiation term appears to act almost like a curvature
term.

\item
We free up the curvature of space, but exclude dark radiation, \ie
$\kappa \neq 0, C = 0$.

In this case, the reduced Hubble parameter is given by
\ber\label{eq:hubble2}
h^2(z) &=&  \omr(1+z)^4+\omt(1+z)^3+\omk(1+z)^2+\omsig+2\oml \\\nonumber 
&&-2\sqrt{\oml}\sqrt{\omr(1+z)^4+\omt(1+z)^3+\omsig+\oml+\omlb}\,\,,
\eer 
with the additional constraint relation
\beq\label{eq:omsig2}
\omsig=1-\omr-\omt-\omk +2\sqrt{\oml}\sqrt{1+\omlb-\omk}\,\,. 
\eeq

The effective equation of state at present is now given by 
\beq
w_0 = -1 -\frac{1}{3}\frac{\sqrt{\oml}(4\omr+3\omt)}{(1-\omr-\omt-\omk)(\sqrt{1+\omlb-\omk}+\sqrt{\oml})}\,\,.
\eeq

The parameters to be fitted are $\omt, \ \oml, \ \omlb, \ \omk$ and
$H_0$. Current CMB measurements show that the universe is practically
flat, with $\omk \sim 0$, for the cosmological constant, as we shall
see this strong constraint may not hold in the braneworld scenario.

\end{itemize}

It is possible to consider a model including both the dark radiation
and curvature terms, but since both terms have a similar effect on the
expansion of the universe (both being proportional to $\sim (1+z)^2$),
we expect them to be somewhat degenerate with each other, so it would
not be possible to easily discriminate them using distance measures
alone.

\section{Data and Methodology}\label{sec:data}

We use here the cosmological data that gives quasi-model-independent
information on the background expansion of the Universe. The most
commonly used data for this purpose is the Supernova Type Ia
\cite{union,jla}. There are also the Baryon Acoustic
Oscillations\cite{bao,baoh, sdss12}, the comoving size of the sound
horizon at last scattering surface from CMB data \cite{planck}, the
value of Hubble parameter derived from various independent sources
\cite{h_var}, Gamma Ray Bursts \cite{grb}, direct measurements of the
Hubble constant $H_0$ \cite{riess_h0, efst_h0, riess16_h0} \etc.

Not all the data is regarded with the same degree of confidence, \eg
the Gamma Ray Bursts observations meet with some scepticism from the
community due to the large scatter in their intrinsic properties. We
therefore choose not to utilize these data in our analysis.

Direct measurements of $H_0$ are also subject to various tensions. The
HST Cepheid+SNe based estimate from \cite{riess_h0} gives $H_0 = (73.8
\pm 2.4) \ {\rm km} \ {\rm s}^{- 1} \ {\rm Mpc}^{-1}$. The same
Cepheid data have been re-analysed in \cite{efst_h0} using revised
geometric maser distance to NGC 4258. Using NGC 4258 as a distance
anchor, they find $H_0 = (70.6 \pm 3.3) \ {\rm km} \ {\rm s}^{-1}
\ {\rm Mpc}^{-1}$. A recent paper, \cite{riess16_h0}, obtains a
$2.4\%$ determination of the Hubble Constant at $H_0 = 73.24 \pm 1.74
\ {\rm km} \ {\rm s}^{-1} \ {\rm Mpc}^{-1}$ combining the anchor NGC
4258, Milky Way and LMC Cepheids. This value disagrees at $3\sigma$
with that predicted by Planck for the $\Lambda$CDM 3-neutrino model in
\cite{planck}, which is $H_0 = 67.3 \pm 1.0 \ {\rm km} \ {\rm s}^{-1}
\ {\rm Mpc}^{-1}$. The Milky Way Cepheid solutions for $H_0$ may be
unstable \cite{efst_h0}, which could go some way in explaining this
inconsistency. Recent strong lensing observations, \cite{bonvin16_h0},
give the value $H_0 = 71.9^{+2.4}_{-3.0} \ {\rm km} \ {\rm s}^{-1}
\ {\rm Mpc}^{-1}$. On the other hand, the Planck results appear to
favour a lower value of $H_0$ \cite{planck}. Hubble parameter
measurements from SNe and red giant halo populations, \cite{tamm13_h0}
give $H_0 = 63.7 \pm 2.3 \ {\rm km} \ {\rm s}^{-1} \ {\rm
  Mpc}^{-1}$. A recent Hubble parameter measurement by
\cite{ratra16_h0} prefers a value of $H_0 = 68.3^{+2.7}_{-2.6} \ {\rm
  km} \ {\rm s}^{-1} \ {\rm Mpc}^{-1}$. The most recent SDSS DR12 BAO
data \cite{sdss12} also appears to favour a somewhat low value of $H_0
= 67.8 \pm 1.2 \ {\rm km} \ {\rm s}^{-1} \ {\rm
  Mpc}^{-1}$. Historically, direct measurements of $H_0$ have often
resulted in widely discrepant values. Even today, some measurements
find comparatively higher values of $H_0$ than others. There are also
issues with the reliability of analysis for the different datasets. In
our analysis, we do not use any priors on $H_0$ and let the analysis
choose the preferred value of $H_0$.

The cosmic chronometer datasets, which estimate the Hubble parameter
with different evolution of cosmic chronometers in the redshift range
$0<z<2$ have been recently used in \cite{stat} to constrain the
equation of state. These datasets may be somewhat dependent on the
assumptions of evolutionary stellar population synthesis models, they
also rely on the correct identification of tracers and reliable age
dating. The constraints obtained from these datasets in conjunction
with other data appear to favour phantom behaviour over $w>-1$ models,
therefore these datasets may well fit our Phantom braneworld models
successfully. For the moment we leave this dataset out, since the
assumption dependence of these datasets is still being studied.

We create here a base dataset comprising of those observations whose
systematics are well constrained, or which have already been used with
some success in conjunction with each other.

\subsection{Supernova Data}

We use the Union2.1 SNe Ia dataset \cite{union} comprising of $580$
SNe between $z \sim 0.01-1.4$, with average errors $\sigma_{\mu} \sim
0.1-0.6$. One can also use the JLA dataset \cite{jla} which combines
the SNLS and SDSS SNe to create an extended sample of 740 SNe, with
apparently better calibration quality, but this does not appreciably
change results. We use the full SNe error covariance matrices for the
analysis. The data is in the form:
\beq
\mu(z) = 5 \ {\rm log}_{10} \left( \frac{c(1+z)}{H_0} \int_0^z \frac{dz}{h(z)}\right) \,\,,
\eeq
with $h(z)$ given by Eq~(\ref{eq:hubble}). It should be noted that at
the redshifts considered, the radiation density is negligible, and
that the only effect of the parameter $H_0$ is as an additive
constant. Thus marginalizing over $H_0$ does not affect the SNe
results.

\begin{table*}
\caption{\footnotesize 
BAO data from different surveys. The two high $z$ \lya points have a
distinct character to the low redshift data, and the data are often
divided into two sets-- low redshift Galaxy BAO data and high redshift
\lya data.}
\label{tab:bao}
\begin{tabular}{cccccccc}
\hline
Source&$z$&$D_V/r_d$&$\sigma$&$D_M/r_d$&$\sigma$&$D_H/r_d$&$\sigma$\\
\hline
6dFGS&$0.106$&$3.047$&$0.137$&$--$&$--$&$--$&$--$\\
SDSS-MGS&$0.15$&$4.480$&$0.168$&$--$&$--$&$--$&$--$\\ 
BOSS-LOWz&$0.32$&$8.594$&$0.095$&$8.774$&$0.142$&$25.89$&$0.76$\\   
BOSS-CMASS&$0.57$&$13.757$&$0.142$&$14.745$&$0.237$&$21.02$&$0.52$\\
\hline
LyaFauto&$2.34$&$--$&$--$&$37.675$&$2.171$&$9.18$&$0.28$\\
LyaF-QSOcross&$2.36$&$--$&$--$&$36.288$&$1.344$&$9.00$&$0.30$\\
\hline
\end{tabular}
\end{table*}

\begin{figure}
\begin{center}
\includegraphics[width=0.4 \linewidth]{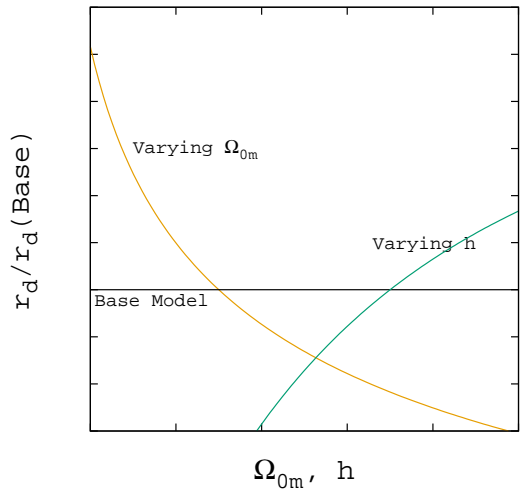} \\ [0.0cm]
\caption{\footnotesize 
Variation of $r_d$ with $h, \omt$. The black line represents the Base
Model with $\omt =0.3, \ h=0.7, \ \oml=\omlb=0$, the orange line
represents the variation of $r_d$ with $\omt$, and the green line
represents the variation of $r_d$ with $h$.}
\label{fig:rdrag}
\end{center}
\end{figure}

\subsection{BAO data}\label{sec:bao}

The current BAO data may be divided into the low redshift Galaxy BAO
data, and the higher redshift \lya data (See table~\ref{tab:bao},
following \cite{bao},\cite{sdss12}). The low redshift data typically
measure a combination of the angular diameter distance and the Hubble
parameter, while the BOSS survey is able to get separate
measurements on both the angular diameter distance and the Hubble
parameter. For the Galaxy data, we use the latest SDSS $12^{th}$ Data
Release \cite{sdss12}, while for the high redshift \lya data we use
the SDSS $11^{th}$ data release \cite{baoh}, since the $12^{th}$
release is not yet available for these. In their most
model-independent form, the observations are presented as a ratio of
between the distance measure ($D_M, D_H, D_V$) and the quantity $r_d$,
which is the comoving sound horizon at the end of the baryon drag
epoch. Therefore the quantities measured model independently are
$D_V/r_d, D_M/r_d, D_H/r_d$, which are given by:
\ber
r_d &=& \frac{1}{H_0} \int_{z_d}^{\infty} \frac{c_s(z)dz}{h(z)}; \ \ c_s(z)=\frac{c}{\sqrt{3}\sqrt{1+0.75\frac{\omb h^2}{\omg h^2(1+z)}}} \\
D_H(z) &=& \frac{c}{H_0 h(z)} \\
D_M(z) &=& \frac{c}{H_0} \int_0^z \frac{dz}{h(z)} \\
D_V(z) &=& [z D_H(z) D_M^2(z)]^{1/3} \,\,.
\eer
where $h=H_0/100$, $\omb$ is the baryon energy density, $\omg$ is the
photon energy density. Typically, for the observations where $D_M,
D_H$ are available separately, we use these directly, taking into
account the covariance between them. Where separately measurements are
not available (6dFGS, SDSS-MGS), we use the combination of these two,
\ie $D_V$.

There are two points to note in the above equations. Firstly, in
$r_d$, we have the sound speed $c_s(z)$ which depends on the ratio of
baryon energy density and photon energy density. We may input $\omb
h^2$ from BBN considerations and $\omg h^2$ from CMB temperature using
the standard scenario, which are both independent of braneworld
parameters or other cosmological parameters except the radiation era
physics.

Secondly, note that, due to the ratios taken, the quantity $h=H_0/100$
does not appear as a multiplicative or additive in the BAO data. It
only appears inside $h(z)$, as part of the radiation term, since the
CMB constraint on this term is on $\omg h^2$, rather than on
$\omg$. For all the quantities at low $z$, the effect of the radiation
term is negligible, as in the SNe data, however, for the drag
distance, $r_d$, it will be significant and neglecting the radiation
term for $r_d$ will lead to erroneous results. One can assume the
$r_d$ obtained from Planck, or use an approximation for it, however,
since these are usually obtained for $\Lambda$CDM with typical values
of $\omt, h$ \etc, so in an analysis where both $\omt$ and $h$ are
parameters, this could change/bias the results by several percent. See
fig~\ref{fig:rdrag} for some illustrative examples of the variation in
$r_d$ with $\omt$ and $h$ (The braneworld parameters are not relevant
at these early times). Therefore the correct way to deal with this
term is to calculate it analytically at each step, for each value of
$\omt$, and marginalizing over $h$. We assume the Planck value for the
drag redshift $z_d=1059.68$ for this, as we do not expect that $z_d$
is as sensitive to the cosmology as $r_d$.

We also note here that, when interpreting the BAO results in the
framework of braneworlds, we implicitly assume that the acoustic sound
in the baryon-photon plasma propagates until recombination with the
same speed as in general relativity.  This assumption holds as long as
the brane effects are negligible during homogeneous cosmological
evolution prior to recombination. Since recombination occurs at high
redshift, we expect that all possible brane effects on the BAO prior
to recombination can safely be neglected. A comparison of results
obtained from the BAO and from the matter power spectrum data for
similar surveys using a self-consistent perturbation theory for the
braneworlds would give us a good handle on the brane effects prior to
recombination.

\subsection{CMB data}\label{sec:cmb}

It is often the practice in cosmological circles to reduce the full
CMB likelihood information to a few background expansion parameters
(\eg as discussed in \cite{cmb1}, \cite{cmb2}). It is possible to
compress a large part of the information contained in the CMB power
spectrum into just a few numbers: specifically the CMB shift parameter
$R$ (\cite{cmb_efst}), and the angular scale of the sound horizon at
last scattering $\la$, dependent on the baryon density
$\omb h^2$ and the scalar spectral index $n_s$:
\ber
R = \sqrt{\omt H_0^2} D_A(z_{\star})/c \\
\la = \pi D_A(z_{\star})/r_s(z_{\star}) \,\,,
\eer
where $D_A(z)$ is the comoving angular diameter distance, and $r_s(z)$
the comoving sound horizon at redshift $z$, where $z_{\star}$ is the
redshift for which the optical depth is unity.

The conservative Planck estimates for these quantities are given as
\cite{planckde}: $R = 1.7382 \pm 0.0088; \ \la = 301.63 \pm 0.15$, at
$z_{\star}=1089.9$.  These numbers are effectively observables and
they can be applied to models with either non-zero curvature or a
smooth dark energy component \cite{cmb3}. However, it has been shown
in \cite{cmb4} that the constraints on these quantities, especially on
$R$, are sensitive to changes in the growth of
perturbations. Therefore, as also mentioned in \cite{planckde}, one
needs to be careful when using these parameters on modified gravity
models which are expected to have very different perturbations to the
standard dark energy models. We therefore use these observables, but
also show the results without them for comparison.  We first use the
observables $\la$ and $R$ separately, fixing $z_{\star}$, to see how
they differ. Then for comparison, as in some recent work on modified
gravity (\cite{modgrav_cmb}), we also use the Planck 2015 priors on
$w$CDM cosmology and the full polarization data for these parameters,
which involves the priors $\lbrace \la = 301.787 \pm 0.089, \ R =
1.7492 \pm 0.0049, \ z_{\star} = 1089.99 \pm 0.29 \rbrace$ and the
inverse covariance matrix
\[ C^{-1} = \left( \begin{array}{ccc}
162.48 & -1529.4 & 2.0688 \\
-1529.4 & 207232 & -2866.8 \\
2.0688 & -2866.8 & 53.572 \end{array} \right).\] 

\section{Results}\label{sec:res}

We first study our base braneworld model, namely the spatially flat
Phantom brane model with no dark radiation (\ie $\omk = 0, \omdr = 0$)
using various combinations of the different datasets to determine the
biases in the observations and to determine which combination of the
data to use for the full analysis. In our analyses we find that the
parameter $\omlb$ has negligible effect for all the different
scenarios, indeed the constraints on the other parameters are
practically the same irrespective of the value of $\omlb$ in all
cases. Therefore, although we mention its best-fit and $1\sigma$ error
levels, we do not depict it in any of the figures that follow.

\subsubsection{Low and High $z$ BAO data}

\begin{figure*}
\begin{center}
\includegraphics[width=1.\linewidth]{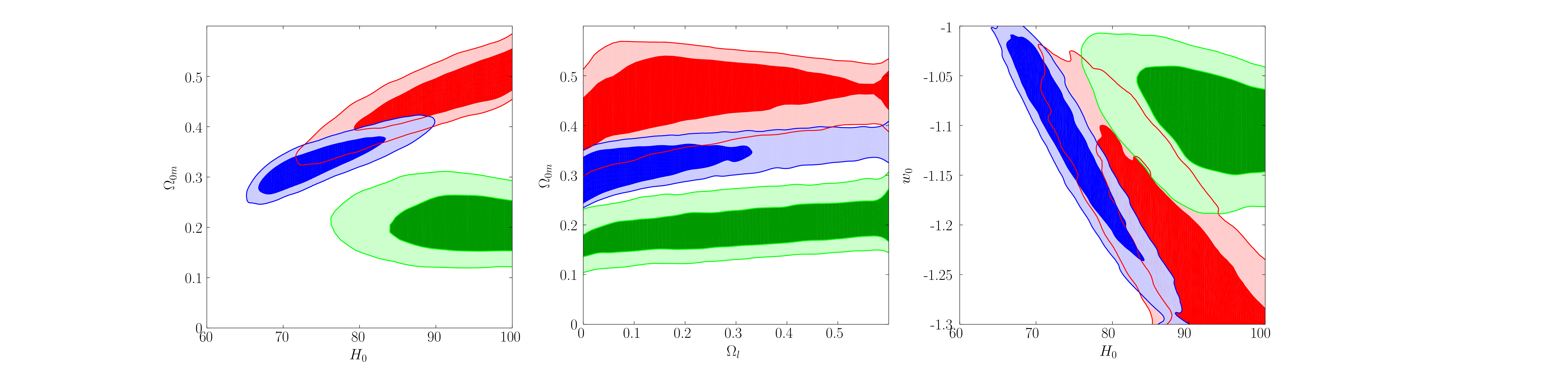} \\ [0.0cm]
\caption{\footnotesize 
$1, 2 \sigma$ confidence levels in the $\omt-H_0$ (left panel),
$\omt-\oml$ (middle panel), $w_0-H_0$ (right panel) parameter space
for the base Phantom brane with $\omk=0, \omdr=0$, using BAO data. The
blue contours represent results for the full BAO data, the red
contours for low $z$ galaxy data only, and the green contours for the
high $z$ \lya data only. The high and low $z$ BAO data are discrepant
at $2\sigma$. $\oml=0$ represents $\Lambda$CDM.}
\label{fig:bao}
\end{center}
\end{figure*}

Unlike the SNe data, the BAO data can be affected by the value of of
the Hubble parameter, due to the effect on $r_d$, as illustrated in
fig~\ref{fig:rdrag}. We attempt to study the effect of $H_0$ on both
high and low redshift BAO data. For low redshift Galaxy BAO data, high
values of $H_0$ lead to correspondingly high values for $\omt$, which
would naturally be ruled out by other observations, while for high
redshift \lya BAO data, high values of $H_0$ lead to slightly lower
values of $\omt$. This obvious discrepancy has also been be noted in
the fig. 4 of \cite{bao} for the $\Lambda$CDM model, and for the older
SDSS DR11 data. We find here that the new DR12 Galaxy data continues
to have the same discrepancy with the \lya data. This has the
interesting consequence that, for the Galaxy BAO data, high values of
$H_0$ are ruled out simply because they would lead to unacceptably
high values of $\omt$, i.e., a high value for the combination $\omt
h^2$, which would come into conflict with most other measurements. But
for the \lya BAO data, even for a high value of $H_0$, the combination
$\omt h^2$ would still be acceptable, and ruling out high values of
$H_0$ would rest on other, more direct observations of $H_0$. This
inconsistency may be due to some systematics in the data itself, or a
true high redshift effect. First reported in \cite{bao} for SDSS DR11,
this apparent discrepancy has also recently been studied in
\cite{bao2} for the same dataset and it has been claimed that the BAO
data at $z > 0.43$ is discrepant with $\Lambda$CDM at $2.8\sigma$. Our
findings for the SDSS DR12 dataset is commensurate with these results
and shows the about $2.3\sigma$ discrepancy between high and low
redshift BAO data. Thus, although somewhat mitigated due to the
degeneracy with braneworld parameters, the disparity that was seen in
the $\Lambda$CDM model is not entirely removed in the braneworld model
either. This then also raises the question whether one should use all
the BAO data available together, or use the Galaxy BAO data and \lya
BAO data separately, since there is clearly some tension between
them. In this paper, we use the entire BAO dataset for final results,
while also showing the results for the Galaxy and \lya data separately
when required. No assumptions or priors are set on the value of $H_0$.

We first check the results for the BAO data for the Phantom brane
scenario, with $\omk=0,\omdr=0$ both at high and low redshifts
separately, and in conjunction. The results are shown
fig~\ref{fig:bao}. We see that both the high and low redshift BAO data
appear to favour higher values of $H_0$ but where the low redshift
data also prefers high values for $\omt$, the high redshift data
favours lower values for $\omt$. When taken together, constraints are
much tighter, and commensurate with other measurements of $\omt, H_0$,
due to the tension between the two datasets which rules out a fair
part of the parameter space. (One also notes a correlation between
$H_0$ and $w_0$, with higher values of $H_0$ being more supportive of
a lower, and more phantom-like, value of $w_0$.). Interestingly, both
the low and high redshift BAO data appear to rule out $w_0 = -1$ at
$2\sigma$ albeit at very high values of $H_0$. When the two datasets
are taken in conjunction, $w_0=-1$ is allowed at $2\sigma$, as the
value of $H_0$ also becomes low for the total dataset. $\oml$ can have
a fairly wide range of values for both datasets, for differing values
of $\omt$. Thus, despite the tension in $H_0$, the constraining power
of the BAO on the braneworld parameters does not change to a large
extent for different subsets of the data. For further analysis, we
shall use the entire BAO dataset, keeping in mind the tension between
the high and low redshift data.

\subsubsection{Compressed CMB data}

\begin{figure*}
\begin{center}
\includegraphics[width=1.\linewidth]{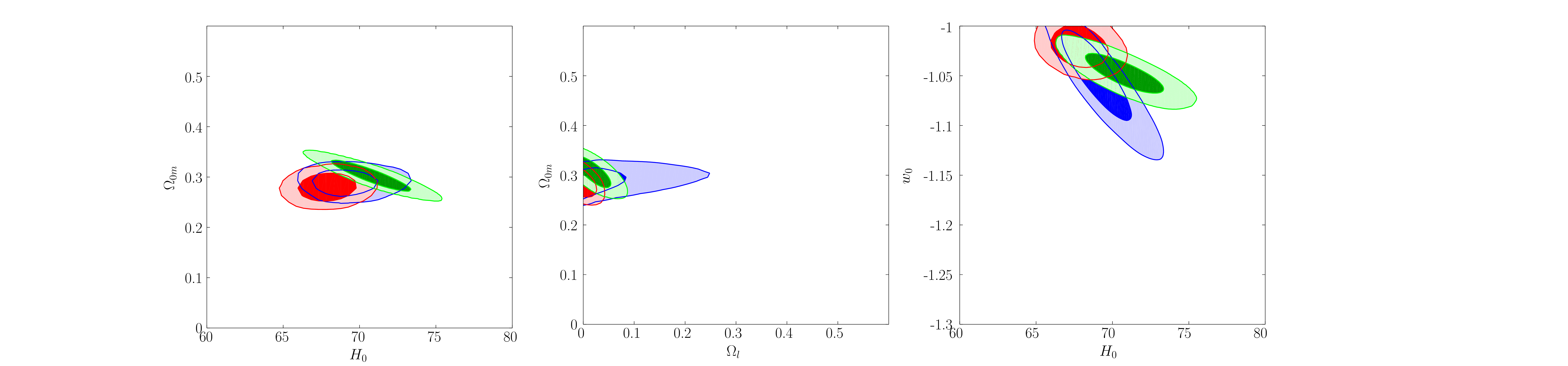} \\ [0.0cm]
\caption{\footnotesize
$1, 2 \sigma$ confidence levels in the $\omt-H_0$ (left panel),
$\omt-\oml$ (middle panel), $w_0-H_0$ (right panel) parameter space
for the base Phantom brane with $\omk=0, \omdr=0$, using compressed
CMB +BAO data. The red contours represent results for the $R$
parameter, blue contours for the $\la$ parameter, and the green
contours for $\lbrace l_A, R, z_{\star} \rbrace$. $\oml=0$ represents
$\Lambda$CDM.}
\label{fig:cmb}
\end{center}
\end{figure*}

We now look at the compressed CMB data for the base Phantom brane
scenario, with $\omk=0,\omdr=0$. We use the BAO data in conjunction
with the CMB since typically a single CMB datapoint is not strong
enough to constrain parameters. We see in fig~\ref{fig:cmb} that the
parameters $R$ and $\la$ give rather different results at $2\sigma$,
with $R$ ruling out a much larger portion of the braneworld parameter
space than $\la$, and also that $R$ prefers slightly lower values of
$H_0$. This also means that $\la$ allows for more negative values of
the effective equation of state today, \ie $w_0 << -1$, as there is a
correlation between higher values of $H_0$ and lower values of
$w_0$. We also see that when using the $\lbrace \la, R, z_{\star}
\rbrace$ dataset, we obtain confidence levels with dgeneracies
entirely different from eiter the $\la$ or $R$ observation, especially
in the $\omt, H_0$ parameter space, \eg this data appears to favour a
larger value of $\omt$ at lower $H_0$. This may be simply a pointer to
the fact that these quantities as derived from standard $w$CDM model
are not compatible with the braneworld models for which perturbations
have not been considered. As has been mentioned in \cite{planck}, the
compressed data is dependent on the perturbations, and so, for
braneworld models which are obviously expected to have very different
perturbations than the standard cosmological constant or scalar field
scenario, the values quoted may not be ideal for use. $R$ can be
especially sensitive to the perturbations. Therefore we do not use the
single observation $R$ in further analysis. We do use $\la$ to better
constrain the degeneracies in the parameters, and we alternatively use
the $\lbrace \la, R, z_{\star} \rbrace$ data, however, we also
simultaneously show the results without the compressed CMB data so
that one can observe the difference made by this CMB.

\subsubsection{Analysis of all datasets}

\begin{figure*}
\begin{center}
\includegraphics[width= 1.\linewidth]{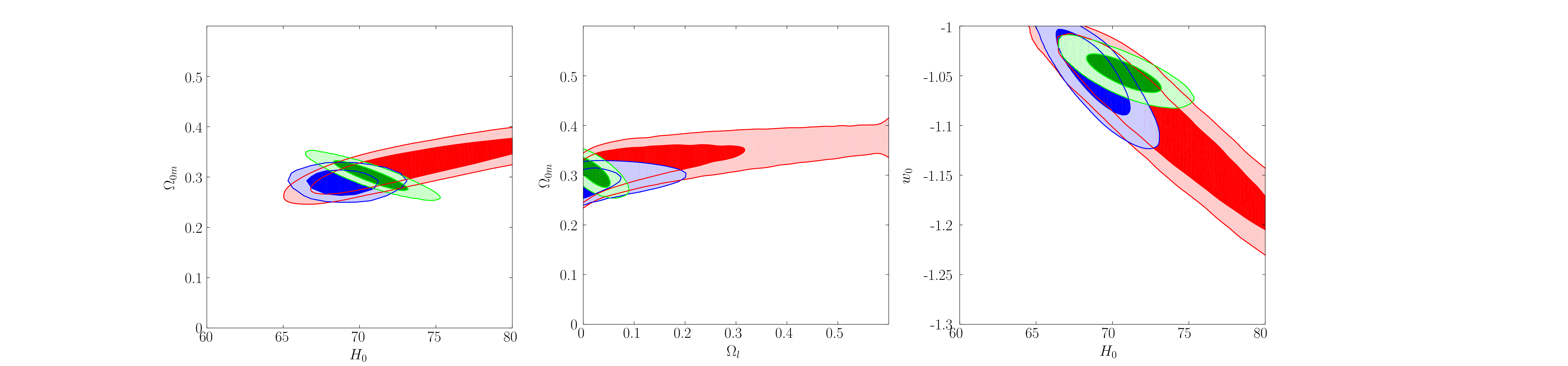} \\ 
\caption{\footnotesize
$1, 2 \sigma$ confidence levels in the $\omt-H_0$ (left panel),
$\omt-\oml$ (middle panel), $w_0-H_0$ (right panel) parameter space
for the base Phantom brane with $\omk=0, \omdr=0$, using SNe Union2.1
+ BAO high and low $z$ data + compressed CMB $\la$ or $\lbrace \la, R,
z_{\star} \rbrace$ data. The red contours represent results for just
the SNe + BAO data, the blue contours use $\la$ in addition, while the
green contours use $\lbrace \la, R, z_{\star} \rbrace$ in
addition. $\oml=0$ represents $\Lambda$CDM.}
\label{fig:B1}
\end{center}
\end{figure*}

In our final analysis of all the three Brane scenarios, we now use the
Union 2.1 SNeIa dataset, the CMB $\la$ data (or alternatively the
$\lbrace \la, R, z_{\star} \rbrace$ data), and the full BAO data. For
the base Phantom brane scenario with $\omk=0, \omdr=0$, the results
are shown in fig~\ref{fig:B1}. We see that the presence of the CMB
data severely limits the allowed values of the $\oml$ parameter. At
$1\sigma$, $\oml \sim 0.13$ for the SNe+BAO data, while including the
CMB $\la$ data limits $\oml \sim 0.08$ at $1\sigma$, while the
$\lbrace \la, R, z_{\star} \rbrace$ data gives the constraints $\oml
\lleq 0.05$ at $1\sigma$. The CMB data also puts much tighter
constraints on the $\omt, H_0$ parameters. In absence of CMB, the SNe
data typically does not in effect constrain these parameters well,
while the low redshift BAO data, as shown in the previous section,
favours somewhat higher values of $\omt$ and $H_0$ than would be
allowed by the CMB observations. These values are ruled out when the
CMB datapoint is added, thus tightening the constraints. The higher
the values of $H_0$ allowed, the more the effective equation of state
shows phantom-like behaviour. Thus the constraints without CMB allows
for $w_0 \lleq -1.19$, for $H_0 \lleq 78 \ {\rm km} \ {\rm s}^{-1}
\ {\rm Mpc}^{-1}$ at $1\sigma$, while the addition of CMB $\la$
constrains the effective equation of state to $w_0 \simeq -1.09$ and
the Hubble parameter to $H_0 \lleq 71 \ {\rm km} \ {\rm s}^{-1} \ {\rm
  Mpc}^{-1}$ at $1\sigma$. The CMB $\lbrace \la, R, z_{\star} \rbrace$
data constrains $w_0 \lleq -1.09$ and $H_0 \lleq 72 \ {\rm km} \ {\rm
  s}^{-1} \ {\rm Mpc}^{-1}$ at $1\sigma$.

\begin{figure*}
\begin{center}
\includegraphics[width= 1.\linewidth]{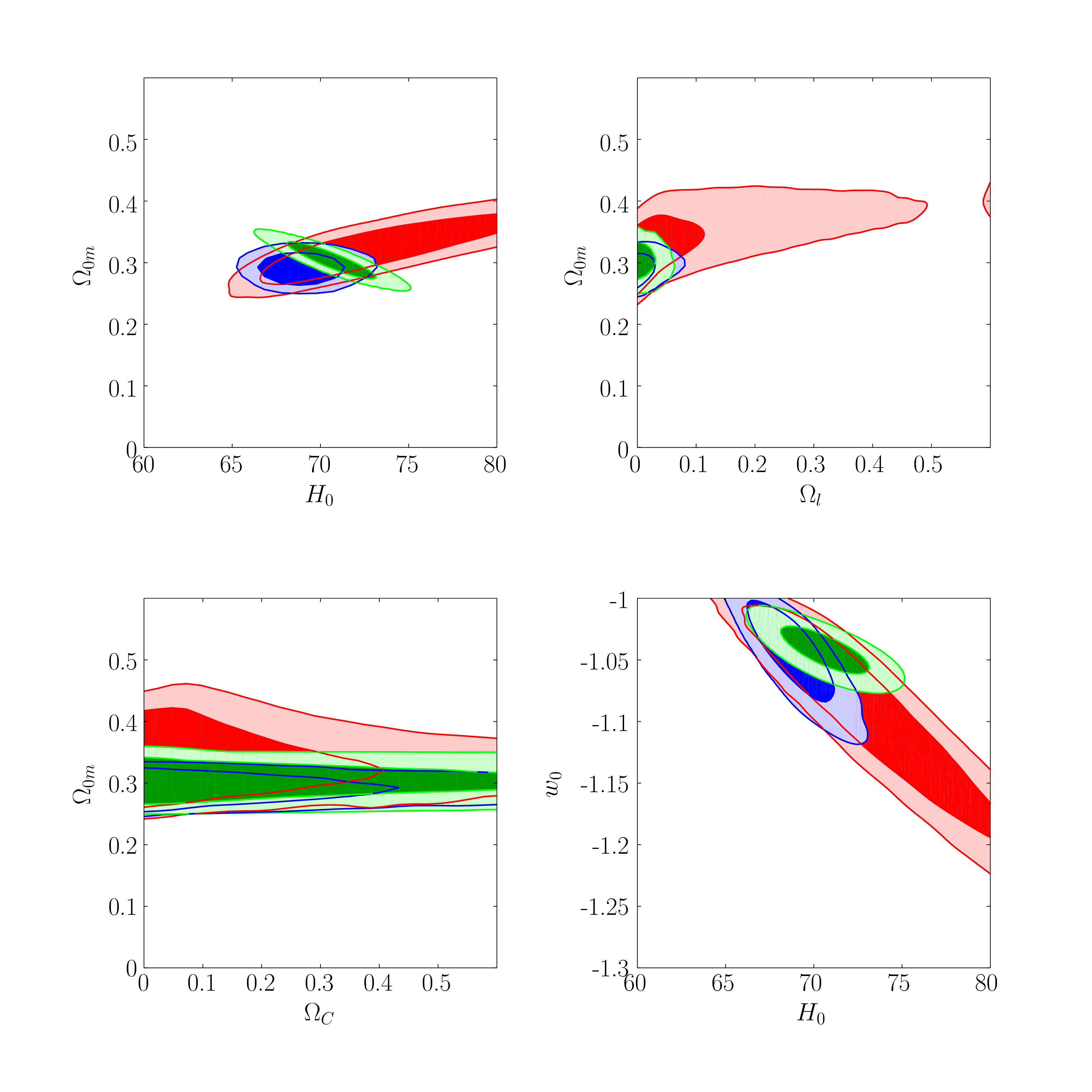} \\ 
\caption{\footnotesize
$1, 2 \sigma$ confidence levels in the $\omt-H_0$ (top left panel),
$\omt-\oml$ (top right panel), $\omt-\omdr$ (bottom left panel),
$w_0-H_0$ (bottom right panel) parameter space for Phantom brane
including dark radiation, using SNe Union2.1 + BAO high and low $z$
data + compressed CMB $\la$ or $\lbrace \la, R, z_{\star} \rbrace$
data. The red contours represent results for just the SNe + BAO data,
the blue contours use $\la$ in addition, while the green contours use
$\lbrace \la, R, z_{\star} \rbrace$ in addition.  $\oml=0$ represents
$\Lambda$CDM.}
\label{fig:B1DR}
\end{center}
\end{figure*}

For the case where dark radiation is considered, the results are shown
in fig~\ref{fig:B1DR}. In this case we find that the presence of the
added $\omdr$ parameter constrains the $\oml$ parameter quite
strongly, and $\oml$ in this case is smaller than in the previous
case. With CMB data, $\oml \sim 0.04$ at $1\sigma$, while for just the
SNe+BAO data, $\oml \sim 0.13$ at $1\sigma$. This is because the term
$\oml$ is present in two terms in the eq~\ref{eq:hubble}, one positive
and the other negative. The best-fit in the $\omdr=0$ case holds for
some ratio of these two terms. A non-zero $\omdr$ changes this ratio
by increasing the negative, square-rooted term, thus necessitating a
corresponding reduction in $\oml$ to offset this increase. As
previously, the Hubble parameter for the SNe+BAO analysis is allowed
to be as high as $H_0 \lleq 80 \ {\rm km} \ {\rm s}^{-1} \ {\rm
  Mpc}^{-1}$, and the corresponding effective equation of state is
$w_0 \lleq -1.2$ at $1\sigma$. The addition of CMB $\la$ constrains
the parameters to $H_0 \lleq 70 \ {\rm km} \ {\rm s}^{-1} \ {\rm
  Mpc}^{-1}$, $w_0 \lleq -1.1$ at $1\sigma$, while addition of
$\lbrace \la, R, z_{\star} \rbrace$ data gives $w_0 \lleq -1.08$ and
$H_0 \lleq 72 \ {\rm km} \ {\rm s}^{-1} \ {\rm Mpc}^{-1}$ at
$1\sigma$.

\begin{figure*}
\begin{center}
\includegraphics[width= 1.\linewidth]{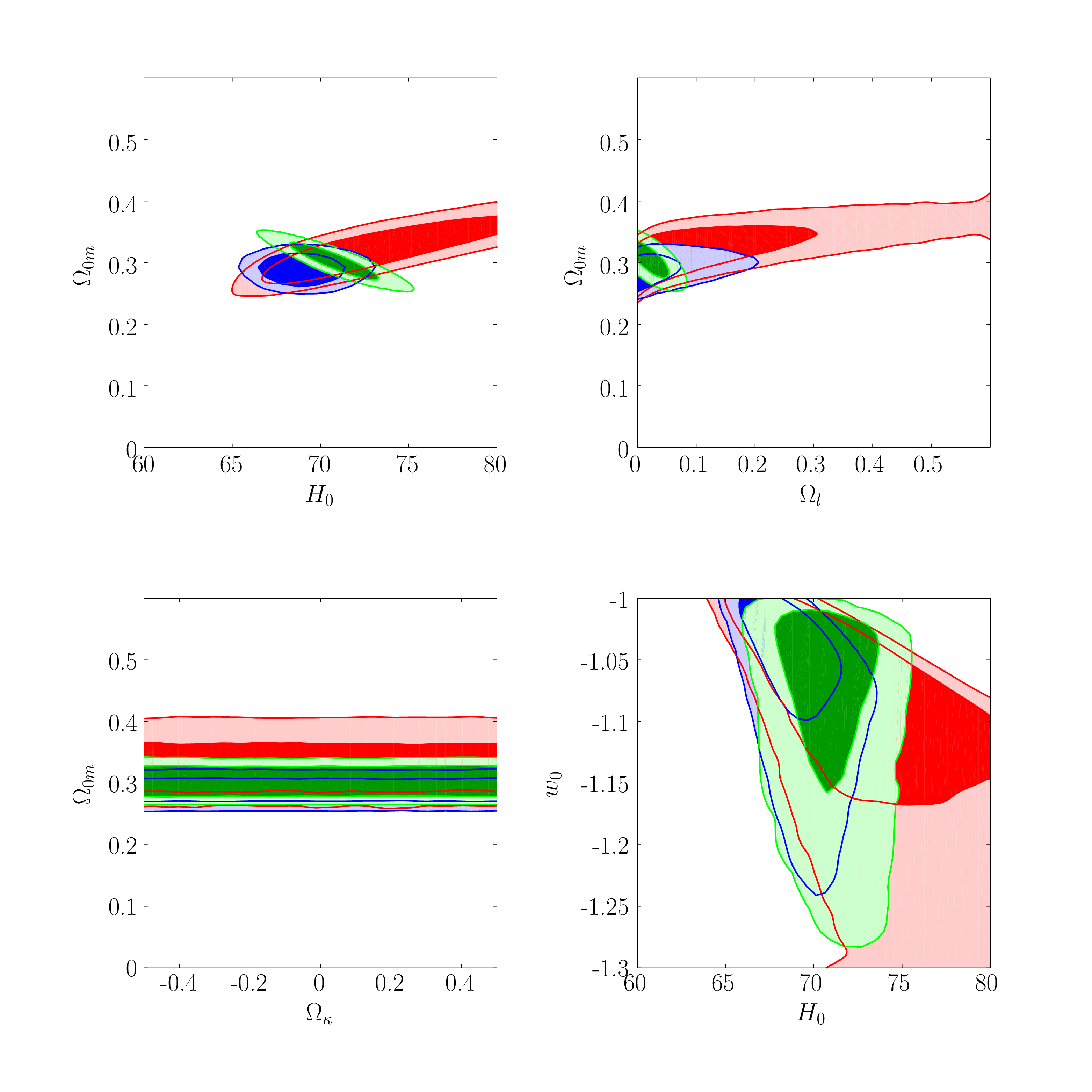} \\ 
\caption{\footnotesize
$1, 2 \sigma$ confidence levels in the $\omt-H_0$ (top left panel),
$\omt-\oml$ (top right panel), $\omt-\omdr$ (bottom left panel),
$w_0-H_0$ (bottom right panel) parameter space for Phantom brane
including curvature, using SNe Union2.1 + BAO high and low $z$ data +
compressed CMB $\la$ or $\lbrace \la, R, z_{\star} \rbrace$ data. The
red contours represent results for just the SNe + BAO data, the blue
contours use $\la$ in addition, while the green contours use $\lbrace
\la, R, z_{\star} \rbrace$ in addition. $\oml=0$ represents
$\Lambda$CDM.}
\label{fig:B1omk}
\end{center}
\end{figure*}

The results for the case where the curvature of the universe is left
as a free parameter are shown in fig~\ref{fig:B1omk}. We find in this
case that the allowed values of $\oml$ for SNe+BAO is roughly the same
as in the first case, $\oml \sim 0.3$ at $1\sigma$, the addition of a
new parameter $\omk$ does not afford much more flexibility in
parameter space. In the case where the CMB is considered, given that
the CMB is expected to constrain the curvature of the universe quite
strongly, $\oml$ is slightly better constrained than the flat case,
with $\oml \sim 0.08$ at $1\sigma$. However, even with these small
values of $\oml$, the curvature of the universe is allowed to be
non-zero, and the universe at $1\sigma$ could either be closed or
open, with $-0.5 \lleq \omk \lleq 0.5$ even when CMB data is
considered. The Hubble parameter is constrained to $H_0 \sim 78 \ {\rm
  km} \ {\rm s}^{-1} \ {\rm Mpc}^{-1}$, and the effective equation of
state to $w_o \sim -1.24$ at $1\sigma$ for SNe+BAO data, and the
addition of CMB $\la$ brings these numbers down to $H_0 \sim 70 \ {\rm
  km} \ {\rm s}^{-1} \ {\rm Mpc}^{-1}$, $w_0 \sim -1.1$. For the
$\lbrace \la, R, z_{\star} \rbrace$ data, constraints are weaker $w_0
\lleq -1.15$ and $H_0 \lleq 73 \ {\rm km} \ {\rm s}^{-1} \ {\rm
  Mpc}^{-1}$ at $1\sigma$.

The table~\ref{tab:bestfit} shows the best-fit and $1\sigma$ errors on
the various parameters $H_0, w_0, \omt, \oml, \omlb, \omdr, \omk$ in
all the cases considered. We note, first of all, that the five
dimensional cosmological constant at $1\sigma$ basically encompasses
its entire parameter space and also that the results are fairly
insensitive to the value of $\omlb$. Thus for most such analyses, we
may neglect the effects of $\omlb$. We note also that without the CMB
data, slightly higher value of $\omt, H_0$, and a lower, more
phantom-like $w_0$, are preferred, and also that the presence of the
CMB data puts quite strong constraints on the $\oml$ parameter which
represents the length scale at which the bulk affects the brane. Using
just the SNe and BAO data, we can constrain $\oml \sim 0.13-0.3$ at
$1\sigma$ for the different models. Including the CMB data brings down
these numbers to $\oml \sim 0.04-0.10$. We also note that, for the
case where the restriction on the curvature of the universe is lifted,
even the inclusion of the CMB data does not appear to rule out closed
or open universes for braneworld models.

We should be cautious, however, about our interpretation of these
results. As we have mentioned in the previous sections, the low and
high $z$ BAO data is discrepant at $2\sigma$, thus results from the
joint analysis of both datasets severely constrains the parameter
space due to the tension between the datasets. Thus the tight
constraints we obtain on the braneworld parameters may very well
change as more BAO data becomes available and this tension between low
and high $z$ data is resolved. We also note that the compressed CMB
data may not be completely appropriate to use for modified gravity
models. Therefore, the correct way to include the CMB in this analysis
would be by doing a complete self-consistent perturbative analysis,
rather than using a single number $\la$ or $R$ or a combination
thereof which has been calculated for the Einsteinian gravity
framework rather than for modified gravity. The severe constraining of
the parameter space thus may be a spurious effect of simply using data
inappropriately.

\begin{table*}
\begin{center}
\caption{\footnotesize 
Bestfit and $1\sigma$ confidence levels on cosmological parameters for
various braneworld models for different datasets.}
\label{tab:bestfit}
\begin{tabular}{cccccccc}
\hline
&$H_0$&$w_0$&$\omt$&$\oml$&$\omlb$&$\omdr$&$\omk$\\
\hline
Phantom brane w $\la$&$69.04^{+2.55}_{-1.42}$&$-1.06^{+0.04}_{-0.03}$&$0.289^{+0.010}_{-0.009}$&$0.047^{+0.031}_{-0.047}$&$0.552^{+0.441}_{-0.552}$&$--$&$--$\\
\\
Phantom brane, $\omdr$ w $\la$&$69.02^{+1.44}_{-1.80}$&$-1.05^{+0.04}_{-0.05}$&$0.291^{+0.016}_{-0.010}$&$0.015^{+0.025}_{-0.015}$&$0.537^{+0.461}_{-0.537}$&$0.253^{+0.147}_{-0.253}$&$--$\\
\\
Phantom brane, $\omk$ w $\la$&$69.06^{+1.42}_{-1.79}$&$-1.07^{+0.09}_{-0.03}$&$0.289^{+0.010}_{-0.009}$&$0.047^{+0.034}_{-0.047}$&$0.552^{+0.439}_{-0.552}$&$--$&$-0.242^{+0.472}_{-0.207}$\\
\\
\hline
\\
Phantom brane w $\lbrace \la, R, z_{\star} \rbrace$&$70.75^{+1.30}_{-1.30}$&$-1.05^{+0.03}_{-0.02}$&$0.303^{+0.011}_{-0.011}$&$0.025^{+0.023}_{-0.025}$&$0.549^{+0.449}_{-0.549}$&$--$&$--$\\
\\
Phantom brane, $\omdr$ w $\lbrace \la, R, z_{\star} \rbrace$&$70.63^{+1.35}_{-1.55}$&$-1.04^{+0.04}_{-0.02}$&$0.304^{+0.013}_{-0.012}$&$0.012^{+0.029}_{-0.012}$&$0.525^{+0.456}_{-0.525}$&$0.265^{+0.315}_{-0.265}$&$--$\\
\\
Phantom brane, $\omk$ w $\lbrace \la, R, z_{\star} \rbrace$&$70.78^{+2.30}_{-1.43}$&$-1.06^{+0.09}_{-0.04}$&$0.302^{+0.012}_{-0.011}$&$0.045^{+0.032}_{-0.025}$&$0.542^{+0.457}_{-0.542}$&$--$&$-0.179^{+0.679}_{-0.321}$\\
\\
\hline
\\
Phantom brane w/o CMB&$75.03^{+3.09}_{-7.11}$&$-1.12^{+0.09}_{-0.07}$&$0.332^{+0.032}_{-0.043}$&$0.222^{+0.077}_{-0.222}$&$0.576^{+0.413}_{-0.576}$&$--$&$--$\\
\\
Phantom brane, $\omdr$ w/o CMB&$75.33^{+4.44}_{-7.67}$&$-1.12^{+0.08}_{-0.08}$&$0.334^{+0.031}_{-0.047}$&$0.098^{+0.031}_{-0.098}$&$0.569^{+0.410}_{-0.569}$&$0.220^{+0.272}_{-0.220}$&$--$\\
\\
Phantom brane, $\omk$ w/o CMB&$74.89^{+3.23}_{-7.05}$&$-1.14^{+0.16}_{-0.10}$&$0.331^{+0.031}_{-0.043}$&$0.218^{+0.091}_{-0.218}$&$0.574^{+0.417}_{-0.574}$&$--$&$-0.185^{+0.423}_{-0.416}$\\
\\
\hline
\end{tabular}
\end{center}
\end{table*}

\section{Conclusions}\label{sec:concl}

In this work, we have used primarily the SNe Type Ia and BAO
observations, as well as compressed CMB data to constrain braneworld
parameters. We find that for the analysis using SNe + BAO data, we are
faced with some tension between low and high redshift BAO
observations, mainly due to their apparently favouring very different
values of the Hubble parameter today. Both datasets considered
jointly, in conjunction with the SNe allow $\oml \lleq 0.3$ at
$1\sigma$ for our base Phantom brane model with $\omdr=0,
\omk=0$. Including the dark radiation term, we find the $1\sigma$
constraint of $\oml \lleq 0.13, \omdr \lleq 0.4$. For the case where
curvature is left to be a free parameter, the results are not very
different for $\oml$, but closed and open universes are allowed at
$1\sigma$, with $-0.5 \lleq \omk \lleq 0.5$. When the compressed CMB
data is added, the constraints become much stronger. For the simplest
case of Phantom brane with $\omdr =0, \omk = 0$, using the CMB
parameter $\la$ the $\oml$ parameter is constrained at $1\sigma$ to
$\oml \lleq 0.1$, for the case with dark radiation, we have $\oml
\lleq 0.04, \omdr \lleq 0.4$, while for the case with non-zero
curvature, we obtain $\oml \lleq 0.08$, while the curvature remains as
unconstrained as just the SNe+BAO data. When utilizing CMB data, the
constraints on the Hubble parameter are naturally very close to the
Planck values for $\Lambda$CDM, while BAO+SNe data by themselves allow
quite higher values for $H_0$ which are more in line with some direct
measurements of $H_0$. Consequently, the effective equation of state
for the SNe+BAO case shows marked phantom-like behaviour, with $w_0
\lleq -1.2$, whereas the addition of CMB constrains it somewhat more,
to $w_0 \lleq -1.1$. We should remember that while the compressed CMB
data is ideally suited for use in the cosmological constant or scalar
field scenarios, it may not be as suitable for modified gravity, which
is expected to have noticeably different perturbations from these
scenarios. Therefore, an analysis of the full CMB data with
self-consistent perturbations may give entirely different results.

In conclusion, we find that Phantom braneworld models are well
constrained by current distance measures but by no means ruled out. It
is possible to construct braneworld models compatible with the current
observations in which brane-specific effects can cause the
acceleration of the cosmological expansion, thus offering a
complementary approach to the dark energy problem. We note the
discrepancy between high and low $z$ BAO data and quote the most
conservative results using both datasets. Analysis with future BAO
data should make it clearer if this inconsistency is in the data
itself, or requires a more fundamental change in the cosmological
modelling of dark energy. Final constraints on such models can only be
obtained if we are able to self-consistently include the perturbative
effects of the braneworld models. We note here that perturbations on
the braneworld are not expected to modify the transfer function to a
great extent, since it is mostly determined by high-$z$ physics which
remains similar to the cosmological constant in our model.  However,
self-consistent perturbations on the brane are expected to affect: (i)
low-$z$ growth rate through $f(z)$ and $\sigma_8$, (ii) the ISW
effect, since $\Phi$ differs from the $\Lambda$CDM value, and (iii)
weak lensing, since $\Phi \neq \Psi$. A companion paper will explore
these issues in further detail.

\section{Acknowledgements}
The authors would like to acknowledge useful discussions with
Yu. Shtanov and A. Viznyuk.  UA was supported in this project by the
``DST Young Scientist Program'' of SERB, India. UA would also like to
thank the CHEMCHAM team at IRAP, Toulouse for the use of the hyperion2
cluster for some of the calculations of this paper.

\end{document}